# Particle and Wave: Developing the Quantum Wave Accompanying a Classical Particle

C. L. Herzenberg

**Abstract**
The relationship between classical and quantum mechanics is explored here in an intuitive manner by the exercise of constructing a wave in association with a classical particle. Using special relativity, the time coordinate in the frame of reference of a moving particle is expressed in terms of the coordinates in the laboratory frame of reference in order to provide an initial spatiotemporal function to work from in initiating the development of a quantum wave. When a property of temporal periodicity is then ascribed to the particle, a provisional spatiotemporal function for a particle travelling at constant velocity manifests itself as an elementary running wave characterized by parameters associated with the moving particle. The potential locations of the associated particle considered in Lorentz transformations suggest that the quantum wave associated with a particle moving at constant velocity would exhibit a uniformly distributed presence throughout space. An initial wave description for bidirectional motion is generated based on an average time coordinate for a combination of oppositely directed elementary running waves, and the resulting spatiotemporal function exhibits wave behavior characteristic of a standing wave. Ascribing directional orientation to the intrinsic periodicity of the particle introduces directional sub-states. For combined states of motion, the relative number of directional sub-states exhibits a sinusoidal distribution as a function of angular separation in orientation. The variation in the relative number of directional sub-states associated with specific spatial phase differences in standing waves leads to spatially varying magnitudes for standing waves. Further analysis of standing waves representing bidirectional motion and their components leads to full mathematical expression for all waves representing free particle motion. A generalization to wave behavior associated with particles subject to force fields enables us to develop a governing differential equation identical in form to the Schrödinger equation.

**Key Words:**
Quantum wave, classical particle, quantum theory, wave function, de Broglie frequency, special relativity, directional sub-states, Schrödinger equation, Lorentz transformation, orientation direction

**Introduction**

Historically, discussions of quantum mechanics involved the duality of the particle and wave characteristics of objects. Here, we attempt to improve our understanding of the relationship between classical and quantum ideas by undertaking examination of a process of developing quantum wave functions from an origin in classical concepts. Starting with a classical particle, how can we introduce and develop an associated wave in a simple, straightforward, and intuitive manner, and what can be learned from this process? And how can we use parallel dialectical processes to clarify the significance of features of quantum mechanics in classical terms, and vice versa?

Our predominant point of view will be a classical perspective. A classical particle is an object described by its localized position and its momentum, whereas a wave is a field or spatiotemporal function described by its values or magnitudes at every point of space and time. We will first consider a classical particle in the context of special relativity, and we will introduce the particle's relativistic time in the laboratory system of coordinates as an initial spatiotemporal function that will provide our first step toward developing an associated quantum wave. Then we will add supplementary features to the properties of the original classical point particle, and examine the implications of these features for the particle's associated wave.

Primary among the transitional features to be allocated to the particle is a temporal periodicity which is taken to be proportional to the particle's mass. This feature brings us a further step forward to creating candidate quantum waves which have frequencies determined by the particle's energy and wavelengths determined by the particle's momentum.

To move on to more complicated quantum waves, it seems to be necessary to accept that a transitional or semi-classical particle is capable of existing simultaneously in more than a single classical particle state, or of sharing simultaneously the characteristics of differing classical states. We engage these superposed particle states by working with an average time as the basis for forming a provisional spatiotemporal function for developing a quantum wave to associate with these superposed particle states.

A key newly introduced idea is the presence of a directionality to be associated with the temporal periodicity. Consideration of the various possible orientations in three dimensional space for the motion associated with the periodicity leads to a consideration of directional sub-states to be associated with each state of translational motion of the particle. Evaluating the relative number of sub-states that form in combined states in turn leads to the appearance of spatially varying magnitudes for the associated waves. From this information we deduce the full mathematical form of wave functions associated with all free particle states.

A generalization from these criteria to waves associated with particles subject to force fields enables us to develop a differential equation governing these waves, and this turns out to correspond to the Schrödinger equation. We go on to consider some of the implications of the assumptions that have led us from a classical particle to its associated quantum wave.

The concepts used and discussed here will be generally familiar to readers, but we hope that the intuitive approach and the process of dialectically and transactionally moving between classical and quantum concepts may be both informative and enjoyable, as well as perhaps providing an option for a somewhat more adventurous pedagogical approach to introducing elementary quantum mechanics than the 'shut up and calculate' or non-interpretation type of approach too often used in the past.

**Associating a spatiotemporal function with a classical particle in uniform motion**

In the hope of better understanding how quantum mechanics relates to pre-quantum or classical physics, it may be helpful to consider how we can start with a classical (unquantized) particle and develop an associated wave (that is, a function of space and time connected with and descriptive of the particle and its motion). How can such a wave be introduced? How can particle properties be connected to wave properties? What criteria or restrictions would be required in order that such a spatiotemporal function would exhibit the properties of a quantum mechanical wave function? And what are the classical concomitants of such a wave? Quantum mechanics exhibits a wave-particle duality; given a classical particle, where can such a wave come from?

A strictly classical particle appears to have no especially obvious immediate connection to a wave that it could be usefully associated with. Since a wave is an explicit function of space and time, we will need to look for a function of space and time that relates in significant ways to the characteristics and behavior of the particle.

To begin, we will examine how we might go about doing this, by addressing the very simplest case of a classical particle in motion, that of a free particle at rest or in uniform constant velocity motion.

We can start by observing that for constant velocity motion, the space and time in two frames of reference or coordinate systems in motion relative to each other are related by the special relativistic Lorentz transformations joining them. If we select as the two frames of reference a frame of reference moving along with the particle, and a frame of reference fixed in the laboratory, then the relationship between these two frames of reference or coordinate systems will incorporate some features relating to and descriptive of the particle's motion.

Use of Lorentz transformations can enable us to relate time and distance as measured in the particle's co-moving frame of reference or coordinate system with time and distance as measured in the laboratory frame of reference, and these relationships will involve parameters that relate to the particle's state of motion.

Thus it appears that a Lorentz transformation could form a basis for introducing a candidate spatiotemporal function to associate with the particle, which is descriptive of the particle's state of motion. In particular, the time as measured in the particle's co-moving frame of reference can be expressed as a function of the temporal and the spatial coordinate values in the laboratory frame of reference, and this time coordinate value associated with the particle's frame of reference can thus provide us with an initial candidate function of space and time in the laboratory frame of reference that we can associate with the particle.

To present this idea more explicitly and in more detail: Following special relativity, we first effectively synchronize clocks in the rest frame of the particle so that this time coordinate would be the same throughout all space. Then we examine this time coordinate in a different frame of reference, the laboratory frame of reference. A Lorentz transformation provides us with the relationship between the time coordinate t' in the particle's co-moving frame of reference with respect to which the particle is at rest, and the time coordinate t and spatial coordinate x in the laboratory frame of reference (Bergmann, 1942):

$$t' = \gamma \, (t - vx/c^2) \qquad (1)$$

Here v is the relative velocity (the velocity of the particle and its co-moving reference frame relative to the laboratory frame of reference) which is directed along the x axis, c is the velocity of light, and γ is the Lorentz factor or relativistic correction factor, $\gamma = 1/(1 - v^2/c^2)^{1/2}$. The time coordinate value in the particle's co-moving frame of reference t', will be introduced as an initial candidate for a spatiotemporal function of x and t to be associated with the classical free particle. Thus, this time coordinate t' will serve as our initial candidate wave in order to proceed with developing a quantum wave associated with the particle.

Next, we observe that if an otherwise classical particle is characterized by an additional property of exhibiting a simple temporal periodicity, the phases at which this periodic behavior takes place can be examined in different frames of reference. Such a particle with its own periodic behavior can as a result be associated automatically with a further developed spatiotemporal function that, as we shall see, turns out to present some of the most familiar features of an elementary quantum mechanical wave function. In particular, for a particle moving in uniform motion with respect to a laboratory frame of reference, special relativity causes such a temporal periodicity in the particle's rest frame to manifest itself also as a spatial periodicity along the direction of motion in the laboratory frame of reference.

To continue a mathematically more explicit discussion: If we form the product of the time coordinate value t' in the particle's co-moving frame of reference with a frequency f that describes the periodicity associated with the particle in its frame of reference, we can obtain an expression for a phase describing the periodic behavior in the co-moving frame of reference. The phase gives us a measure of the number of periods or cycles that have taken place by the time t', and, within an individual cycle, the extent of completion of that cycle or periodicity. We will express the phase in radians rather than cycles by multiplying ft' by 2π so as to obtain:

$$\varphi(t') = 2\pi f t' \qquad (2)$$

Eqn. (2) gives the phase φ(t') of the wave in radians with respect to the origin of coordinates.

The time-dependent phase in the particle's co-moving frame of reference that we obtain in Eqn. (2) tells us that there will be a temporally periodic oscillation or pulsation that has the same value for the phase throughout all of space in the rest frame of the particle. This behavior corresponds to the behavior of a standard quantum wave function for a stationary particle.

Next, we can use the results of the Lorentz transformation between the particle's co-moving frame of reference and the laboratory frame of reference as expressed in Eqn. (1), and insert that into Eqn. (2), so as to obtain an expression for a phase φ(t, x) describing the wave in the laboratory frame of reference:

$$\varphi(t, x) = 2\pi\gamma(ft - fvx/c^2) \qquad (3)$$

The expression for the phase, φ(x,t) in Eqn. (3) now gives us an expression for a spatiotemporal function that is the next step in our construction of a quantum wave in association with a particle.

Note that by starting with the time coordinate in the co-moving frame of the particle and examining its presentation in the laboratory frame of reference, and then incorporating a periodicity, we have effectively introduced a function of x and t that extends throughout all of space and time and that is directly associated with the uniformly moving particle. This function corresponds to the phase associated with the periodicity.

We can deduce from Eqn. (3) that the phase in the laboratory frame of reference represents a traveling wave having phase velocity $c^2/v$, as a fixed value of the phase would move to larger values of x as the time t increases. The wave described by Eqn. (3) represents a phase wave.

So far, we have incorporated only one classical property of the particle into this candidate wave, namely, the particle's velocity. Our introduction of a candidate wave to associate with a classical particle has not yet called on any other prominent characteristic of the classical particle, and in particular does not involve its mass. However, in moving toward a quantum mechanical description, we have instead introduced a separate supplementary transitional property, a periodicity.

How could such a periodicity relate to a classical particle? Since periodic behavior seems to be fundamental to all quantum particles, it would make sense to try to associate this periodicity with a basic property of all classical particles. One of the most basic properties of a classical particle is its mass. A connection between particle mass and quantum frequency has been long recognized, with a history dating to the earliest days of quantum mechanics, when Louis de Broglie introduced the concept of wave-particle duality with the idea that all matter has a wave-like nature, and that every particle would be characterized by an internal periodic phenomenon and would exhibit an associated frequency proportional to its mass or total energy (de Broglie, 1923; Davis; Wikipedia, (1); Wikipedia, (2)).

If we specify that an intrinsic temporal periodicity be associated with the particle, and further require that the frequency should be proportional to the particle's mass, then the spatiotemporal function that we have introduced will turn out to have some of the most prominent and familiar characteristics of a standard quantum wave function. The expression for the phase of this wave will contain a term proportional to the product of the particle's total energy and the time, and also a term proportional to the product of the particle's momentum and the distance along the direction of motion (which leads to a wavelength inversely proportional to the particle's momentum). As a consequence, there will be a phase defined throughout all of space and time, and the function that we have introduced will turn out to describe a traveling wave exhibiting notable similarities to the corresponding quantum wave function for the free particle.

> We can further examine the phase of the particle's periodicity as seen in the laboratory frame of reference as follows. We will take the periodicity associated with the particle to be described by a frequency that is proportional to the particle's mass, as:
>
> $$f = m(c^2/b) \qquad (4)$$
>
> Here, m is the particle's mass and the quantity in parentheses provides a constant of proportionality between the frequency and the mass; and c is the speed of light and b is a constant factor not otherwise specified so far.
>
> Then we can reexpress the phase of the wave in the laboratory frame of reference in Eqn. (3) as:
>
> $$\varphi = 2\pi f t' = 2\pi\gamma(ft - mvx/b) = 2\pi\gamma(ft - x/\lambda) \qquad (5)$$
>
> Eqn. (5) gives the phase for a wave in the laboratory frame of reference that exhibits a frequency $\gamma f$ proportional to the total energy, and a wavelength $\lambda = b/\gamma mv$ that is inversely proportional to the particle's momentum. The relativistic correction factor $\gamma$ is close to unity for nonrelativistic motion, and so can be disregarded in most circumstances that we will consider.
>
> Thus, we have developed a provisional spatiotemporal function that amounts to a running wave with a frequency proportional to the total energy of the particle and a wavelength proportional to the momentum of the particle.

We have succeeded in capturing some of the essential features of an elementary quantum wave function for a free particle by using a very simple approach to developing a wave in association with a classical particle. The results so far are based only on using special relativity together with the premise of the existence of a periodicity in association with the particle that exhibits a frequency proportional to the particle's mass. Thus, we have been able in a rather direct manner to develop a preliminary candidate spatiotemporal function to associate with a classical particle, and this function already shows some of the most prominent features of a standard quantum wave function. The phase for this wave arises as an automatic consequence of the Lorentz transformation laws of special relativity taken together with the assumption of an intrinsic periodicity proportional to the mass of the particle. These results as well as other approaches clearly suggest that quantum mechanics must be to some extent an inherently relativistic phenomenon (French and Taylor, 1978; Davis; Wikipedia (2)) Thus, special relativity appears to be present in an essential manner in quantum mechanics, including in what is referred to as non-relativistic quantum mechanics.

Notice that the spatiotemporal functions that we have introduced are elementary waves that carry phase information only. In essence, so far we have developed a major part (the phase of the wave) of a standard quantum wave function associated with a free particle traveling at constant velocity. We must look further to develop the full mathematical and functional form of such a wave.

**Where is the particle associated with the wave?**

We will take the provisional spatiotemporal function that we have developed that is expressed in Eqn. (5) to provide a partial initial description of a wave that can be associated with and in some sense represent the state of a uniformly moving particle, and then go on to explore what else can be learned in attempting to develop a more complete mathematical description of a quantum wave function to associate with a classical particle.

The spatiotemporal function that we have been able to develop so far in association with a classical particle exhibits features explicitly related to some important particle properties, specifically, the particle's velocity and its mass, - but not that other important characteristic of a classical particle, the particle's spatial position.

The particle's position has not been incorporated into the candidate wave. We have, in effect, introduced a wave in association with the particle that does not refer specifically to the particle's actual position. From our procedure and the form of the provisional spatiotemporal function that we developed it can be seen that this same wave could just as well accompany the particle if the particle were located at any other position in space, as long as its velocity remained the same. Accordingly, if it appears that this wave could be associated with the particle irrespective of the particle's location in space; and therefore this wave could be equally descriptive of any particle having this velocity that could be located anywhere else in space.

The procedure that we have introduced to associate a wave with a classical particle in motion is based on a Lorentz transformation. Lorentz transformations relate the properties of space and time in different reference frames in motion with respect to each other. The Lorentz transformations are transformations between relatively moving frames of reference, and the actual spatial location of the particle itself in its co-moving frame of reference does not enter the Lorentz transformation equation that we have used. Only the velocity of the particle (equal to the velocity of its co-moving frame of reference) relative to the laboratory frame of reference, not the particle's location, enters into the Lorentz transformation. Any other particle anywhere in space that is at rest with respect to the original particle and thus has the same velocity with respect to the laboratory could be equally well characterized by this Lorentz transformation, and any other such particle anywhere in space that also has the same value of the mass would be equally well described by the spatiotemporal (phase) function that we have developed and introduced in Eqn. (5).

So, in the simplest case of a free particle in uniform constant velocity motion, we see that it seems to be possible to associate the same wave with a particle, independently of the particle's position in space. Thus, we would be led to associate such a wave with a particle position anywhere in space, or with a spatially uniform distribution of possible particle positions.

What are the implications of this conclusion that such a wave in association with a particle would not simply be associated with this one particle in particular, but rather might be associated with any similar particle having the same mass and velocity, and located anywhere in space? If we had started with the wave instead of with the particle, any of these particles could equally well be considered the wave's affiliated or associated particle. There is thus evidently not a one-to-one relationship between the particle and the associated wave. Rather, it would appear that in some sense many different incarnations of the particle could equally well be associated simultaneously with such a simple individual wave. Alternatively, we might consider speaking of a uniform distribution of the particle throughout space (or an equal probability of finding the particle anywhere in space) in association with this simplest function to describe the free particle's associated wave. This appears to be telling us, rather simply and clearly, that the particle associated with this wave could be present anywhere in space.

It is interesting and perhaps important to consider the fact that the phase of this wave originates from and is present in the wave on the basis of consideration of any single case of a particle and its associated single Lorentz transformation. However, establishing the particle distribution associated with the quantum wave function would seem to require consideration of many different cases of particle locations and their similar Lorentz transformations for a whole ensemble of particles located at every position in space. Perhaps that could be expressed functionally by considering a combination of many elementary waves or spatiotemporal functions of the type that we have been examining initially, each associated with a particle at a different location. Thus we may need to look at an assembly of particle states and their elementary waves to get the sort of particle distribution described by ordinary quantum wave functions.

Thus we see that a quantum wave associated with a particle moving at constant velocity would be expected to exhibit a uniform distribution of the particle throughout space, as well as being characterized by the phase given in Eqn. (5). The uniformity in distribution of classical localized particle states in association with a wave representing a particle moving at constant velocity should be encoded finally in a full mathematical representation of the wave so that the full wave form will contain information as to the locality or local presence of the particle or particle distribution, as well as the phase information. If we take uniformity in magnitude of the wave as a criterion for uniformity in particle distribution, we should seek a wave of uniform magnitude, which must also incorporate the wave phase.

Introduction of a mathematical representation that provides for a constant magnitude of the wave while simultaneously encoding the spatiotemporal phase relationship would seem to be a way to proceed. This approach would of course lead us structurally but without further derivation or fully satisfactory intuitive justification to the form of a conventional plane wave in complex exponential form, $Ae^{i(\omega t - kx)}$. But rather than simply accepting a structural argument based primarily on mathematics, we will look at further physical aspects of the problem. So as to better understand the origin and role of the mathematical form characterizing an elementary constant velocity quantum wave, we will seek to explore further aspects of the problem that may shed additional light on other aspects of the classical-quantum connection in order to construct a mathematical expression for these waves.

We have noted that considerations of special relativity suggest that there is not a one-to-one relationship between a classical localized particle state and this associated wave, but rather that there would appear to be a many-to-one relationship between classical localized particle states and this wave associated with the particle. Analogously, would many different waves corresponding to different velocities or wave lengths be associated with a single, individual localized particle? And if so, how would these waves combine to characterize such an individual localized particle? More generally, w should examine the question of how quantum waves might combine.

**Initial examination of bidirectional (back-and-forth) motion**

We will next take a look at what might happen when we combine different states of motion, as this may be crucial to developing a deeper understanding of the relationship of classical and quantum descriptions. We will approach this by examining another fairly simple case of particle motion, and see what more we can learn about the characteristics of an associated wave. In this case, we will examine what wave we might be able to associate with a particle moving back and forth at constant speed although not at constant velocity. This is bidirectional linear motion, motion along two opposite directions, or back-and-forth or to-and-fro motion.

In many types of classical motion, we need to address situations in which a particle is capable of moving both forward and backward. In classical physics, these actions are taken to occur sequentially. Quantum mechanics opens the possibility of particle motion occurring both forward and backward simultaneously.

Constant speed motion back-and-forth would appear to be the simplest case of a particle moving both forward and backward, so we will examine this case now. We can initiate examination of the back-and-forth motion of a particle by considering how to combine the features associated with the oppositely directed motions of such a particle. Just as we would combine the separate constant velocity motions of a classical particle in order to attempt to describe back-and-forth particle motion, it would appear useful to consider on what basis information about the waves from the separate motions might lead us to a wave describing a common motion. Using and combining the results on uniform constant velocity motion that we have already developed would appear to offer at least one possible approach to developing a wave associated with back-and-forth particle motion.

How the combination of motions of the particle might take place could be informative. Our original spatio-temporal function associated with a particle engaged in constant velocity motion was introduced on the basis of examining the time coordinate associated with the co-moving reference frame of a moving particle as seen from a different reference frame, the laboratory frame of reference. The oppositely directed motion can be addressed in a similar manner. Then we can attempt a description of the combined motion by performing an average; that is, by evaluating the average value of the relativistic time coordinates associated with the particle motions as seen in a laboratory frame of reference. In forming the average of the time coordinate values, we would weight the time coordinate values associated with the forward and backward motions equally so as to account for the particle participating to an equal extent in the two motions (spending equal amounts of time moving in either direction). As we shall see, the average time then simply turns out to be equal to the laboratory time (upped by a usually small correction from the relativistic γ factor), and this average time turns out to be the same everywhere, throughout all of space, independent of location.

> If we designate the relativistic time coordinate values in the frames of reference associated with the particle motion to the right and to the left as $t'_+$ and $t'_-$ we can, using the appropriate Lorentz transformations, express these time coordinate values in terms of the time and space coordinate values in the laboratory frame of reference as:
>
> $$t'_+ = \gamma (t - vx/c^2) \qquad (6)$$
>
> $$t'_- = \gamma (t + vx/c^2) \qquad (7)$$
>
> Then, using equal weightings for the two separate types of motion, we find for the average particle time:
>
> $$t'_{avg} = \tfrac{1}{2}[t'_+ + t'_-] = \gamma t \qquad (8)$$
>
> Thus, the average time characterizing the back-and-forth motion is very simply related to the laboratory time, becoming equal to the laboratory time in the non-relativistic limit. Like the laboratory time, the average time characterizing the particle motion to- and fro- in the laboratory frame of reference is the same everywhere throughout space.

We should note that the forward and backward motions of the two reference frames are coordinated in the sense that in each case the origins of the two moving coordinate systems associated with particle motion at velocities v and –v have been selected to pass through the origin of the laboratory coordinate system so that all have a common zero time, which we can regard as a time at which a periodic event occurs in all three frames of reference. Thus, the periodicities associated with the incarnations of the particle moving forward and moving backward are coordinated by sharing an event at the synchronized common origin of coordinates. The above particularly simple results are obtained in part because the coordinate systems in the laboratory and in association with the motions forward and backward have been selected to have this particular relationship with each other.

Our preliminary spatiotemporal function describing bidirectional motion is given in Eqn. (8). Our justification for forming a linear combination of waves ultimately is based on the fact that we are forming an average time in order to attempt to describe a physical situation for which no common time is available.

Now that we have evaluated the relativistic time coordinate values for the forward particle motion and for the backward particle motion and for their averaged combination as seen in the laboratory frame of reference, we can look at the corresponding phases. We have seen that in bidirectional motion, as laboratory time changes, the average time changes (although at a slightly different rate), as is evident from Eqn. (8). We can use the behavior of the average time associated with the back-and-forth motion to introduce a phase in association with the average time, as was done earlier for forward motion of a particle using the particle time, in Eqn. (3). At any given laboratory time, this phase for bidirectional motion turns out to be the same throughout all space. If we evaluate the phase associated with the average time in bidirectional motion, it turns out to correspond to the product of the particle energy with the time in the laboratory frame of reference. Thus this wave associated with a particle in back-and-forth motion would exhibit the same temporal periodic phase everywhere in space. Accordingly, this wave associated with back-and-forth motion could be regarded as a standing wave vibrating in phase throughout all of space.

In somewhat more detail: We can introduce phases associated with the particle motion to the right and to the left as $\varphi_+$ and $\varphi_-$. Using Eqn. (5) for the forward motion and introducing a corresponding equation based on Eqn. (7) for the oppositely directed backward motion, we can write the phases as seen in the laboratory frame of reference as:

$$\varphi_+ = 2\pi\gamma \, (ft - mvx/b) = 2\pi\gamma \, (ft - x/\lambda) \qquad (9)$$

$$\varphi_- = 2\pi\gamma \, (ft + mvx/b) = 2\pi\gamma \, (ft + x/\lambda) \qquad (10)$$

In both of these equations, the quantity b/mv has the role of a wavelength and is designated λ. (Actually, the quantity b/γmv is a wavelength also but we will be using the simpler non-relativistic form.)

Thus, the phases of the individual waves have both temporal and spatial components.

However, the phase associated with the average time would, as in Eqn. (2), be given by the product of 2πf with the value of the average time given in Eqn. (8), and would therefore equal:

$$\varphi_{avgt} = 2\pi\gamma ft \qquad (11)$$

We find that the phase associated with the average time for bidirectional motion depends only on the time, and is independent of the spatial coordinate.

Alternatively, if we had averaged the phases given in Eqn. (9) and Eqn. (10), we would obtain a similar result.

But let's return to a point earlier in our examination of back-and-forth motion, specifically to our combining and averaging of the relativistic time coordinate values so as to obtain an average value for the time. We would lose some of the original information relating to the two states of motion if we only combined the two separate time coordinate values into a single averaged time parameter based on the sum of the two separate values. In order to retain the otherwise lost information, we can form a second parameter in terms of the difference between the time coordinate values for the two oppositely directed motions in the laboratory frame of reference.

From these two parameters based respectively on the sum and on the difference of the original time values, the two original relativistic time coordinate values can be reconstructed, so we will not have lost information if we retain both parameters. By using the second parameter, the difference function, we can retain and use all of the additional original information about the bidirectional motion, beyond that already expressed in the average time parameter already introduced. Interestingly, the difference between the relativistic times associated with the back-and-forth motions turns out to give a time-independent quantity proportional to the distance along the direction of common motion.

Previously we combined the time coordinate parameters in Eqn. (6) and Eqn. (7) by addition in order to form an average time, and now we will combine them again, but in subtraction, to form a difference function for the time coordinate values in the laboratory frame of reference:

$$t'_{diff} = -\tfrac{1}{2}[t'_+ - t'_-] = \gamma vx/c^2 \qquad (12)$$

(We have introduced this difference function for the time coordinates as half the difference between their time values, so as to present this companion parameter in a similar format to the average time introduced in Eqn. (8) which was given by half the sum of the two time values.)

We note that this relativistic time difference exhibits no dependence at all on the time in the laboratory frame of reference, and depends only on the distance along the spatial axis of relative motion in the laboratory frame of reference.

Furthermore, we can also examine the spatial behavior associated with the overall back-and-forth motion in terms again of a phase difference (or angular difference between the periodicities) that can be introduced in association with the difference between the relativistic time coordinate values associated with the forward motion and the backward motion, or the corresponding difference beween the phases. It turns out that the corresponding difference between the phases associated with the forward and backward motions is equal to a quantity proportional to the product of the absolute value of the momentum and the distance along the direction of common motion. Accordingly this parameter associated with the combined wave describes a periodicity in space that is time-independent.

Thus, when we combine Eqn. (9) and Eqn. (10) to form a difference function, we find that it corresponds to a difference in phase angle between the two participating waves, and that this is expressed as a spatial dependence:

$$\varphi_{dif} = -\tfrac{1}{2}[\varphi_+ - \varphi_-] = 2\pi\gamma mvx/b = 2\pi\gamma x/\lambda \qquad (13)$$

This difference in phase between the two participating states of motion of the particle originates because the times (and angular phases) associated with the particle in its two different states are different for different values of x in the laboratory frame of reference.

This result can be viewed in terms of a partitioning of the x coordinate segmentally into regions corresponding to the wavelength, in correspondence with the periodic spatial intervals that are observed in conjunction with the two separate directions of motion.

This difference function between the two phases (or angular difference between the two periodicities) is proportional to the distance and is completely independent of the time. As we move along the x-axis, the phases of the two component waves will differ at different values of x. (The two component waves will only be in phase at particular locations of x corresponding to repetitive values of the periodicity, that is, at values of x at intervals corresponding to the wavelength given by $\lambda = b/\gamma mv$.) As x is varied, the spatial phase is increasing for one component wave as it decreases for the other.

This tells us that the spatial behavior of the standing wave depends on the difference in phase (or angular difference between the periodicities) between the two participating states of motion, a fact that we will find useful in subsequent analysis. It will be important to bear in mind that the spatial behavior of the overall combined wave depends on the difference between the phases (or angular difference between the periodicities) associated with the periodicities of the two participating waves.

To summarize, for the case of back-and-forth motion, there will be a common average time throughout all space, and this average time will be simply related to the time associated with a particle at rest, and equal to it in the non-relativistic limit. The net phase associated with the particle's temporal periodicity will also be the same everywhere throughout space. And our results also tell us that every position along the x axis in the laboratory frame of reference represents a particular phase difference between the particle's manifestations of periodicity associated with its two directions of translational motion, and this phase difference at any location is independent of the time.

Thus, the temporal and spatial behaviors turn out to be disjoint for the case of back-and-forth motion. Presumably, a complete spatiotemporal function for bidirectional motion could be mathematically described using a product of separate functions describing the behavior in time and the behavior in space, and would thus describe a standing wave.

So far we have developed some initial spatiotemporal functions with some but not all of the important features of quantum wave functions. Also, so far, we have found only a suggestive basis for a fully detailed mathematical specification of these waves. Where could additional information relevant to such features come from?

**Where is a particle located during bidirectional motion? – an initial inquiry**

Our earlier examination of the wave associated with a particle in constant velocity motion relied on special relativity to present a case that such a free particle would be characterized by a uniformly distributed presence throughout space in the accompanying wave.

But special relativity is only applicable to relative motion at constant velocity, so such an argument is not directly applicable to and not valid for other types of motion. Hence, we must seek an alternative basis for understanding the particle distribution in waves associated with other types of motion.

We know from quantum theory that, unlike the case of a particle in constant velocity motion, the wave function for a particle in constant speed motion to-and-fro does not exhibit a uniform distribution in space. But why is this happening? Why would a particle spend more time (so to speak) in some locations than others? Why should it favor some locations in space? What is affecting the spatial distribution associated with the free particle?

At this point, it appears that we may need to consider further physical phenomena in order to arrive at insight into the spatial location distribution associated with a semi-classical particle. And, in semi-classical terms, perhaps we should also ask why combining separate particle states should affect the particle's probability distribution in space?

It seems to be a fairly general feature of quantum theory that the degree of localization increases with the number of states that combine to form a wave. (This corresponds to the process of building wave packets from component waves). Perhaps we must take into account some additional physical phenomenon that might result in increasing the number of states or sub-states that may exist, and that were insufficiently accounted for in our preliminary consideration of the original semi-classical particle. Perhaps the presence of additional sub-states that might behave differently at different spatial locations might contribute to the non-uniform distribution of particles in most quantum mechanical states, and in particular might contribute to the non-uniform particle distribution in the case of a particle engaged in bidirectional motion.

We will therefore inquire into what further aspects of classical particle behavior that might be relevant to quantum behavior may not yet have been adequately examined, in the expectation that some of these might produce sub-states and thus possibly affect the particle distribution or wave magnitude. We will therefore examine further what we may be able to learn from semi-classical particle states, and attempt to find out what semi-classical effects might affect the magnitude of the wave.

Bearing in mind that we have found that temporal periodicity has already been very helpful in leading to quantum wave properties from a classical starting point, we will begin by considering another possible aspect of periodicity that has not yet been brought into play, and that is the spatial aspect of periodic behavior. Presumably, if an intrinsic periodicity in time present, an associated intrinsic periodic motion must also exist. Could such an intrinsic periodic motion introduce some new considerations based on additional aspects of motion in three dimensional space that might have a bearing on the particle distribution?

**Representing periodicity: what intrinsic motions could be associated with a particle's temporally periodic behavior?**

We can undertake a deeper examination of periodicity by considering the possibility that a spatial manifestation of intrinsic periodic behavior could be present in association with the temporal periodicity.

During the early days of quantum mechanics, Louis de Broglie postulated the existence of the de Broglie frequency, and introduced it without a specified cause. But we will attempt to add an associated physical motion. Let us examine how it might be possible to interpret a de Broglie frequency in terms of a potential repetitive motion of a classical particle. We will inquire how a de Broglie-type temporal periodicity might originate from and be manifested in spatial periodic motion intrinsically associated with the particle. And, further, we will examine how such an intrinsic spatial periodic motion of the particle might affect the waves associated with the particle, and thus might have useful explanatory power for clarifying classical connections to quantum behavior.

What type of motion could be the source of this periodicity? A classical particle, such as the one that we started with, can always be characterized by a mass, a position, and a momentum; and may be characterized by other properties as well. In the simplest case, such a classical point particle would exhibit no periodic behavior. However, there are many sorts of periodic or repetitive behavior in classical physics, and we can look into what types of periodic behavior might be associated with a semi-classical particle as an intrinsic characteristic that might have relevance for developing its quantum behavior.

What spatial behavior of a particle might be associated with temporal periodicity? In principle, there would appear to be a variety of possibilities for periodic or quasi-periodic motion. These could include rotation, vibrational motion, and segmentally continuous stochastic motion, for example. However, rather than examining in detail a particular type of periodic motion, we will confine our attention largely to some general features of periodic behavior and prototype examples of periodic behavior.

**Directional characteristics of periodic motion**

As we have observed, it may be worth looking into some general characteristics of periodic motion to more clearly understand the role of periodicity in quantum mechanics.

One very general feature of any motion, periodic or otherwise, is that a direction or set of directions can be associated with the motion. In particular, any relatively simple periodic or cyclic or oscillatory or otherwise repetitive motion in three dimensional space requires some specification of direction. This might be the axis of a rotational motion, or a direction aligned or anitialigned with vibrational motion, or some other direction associated with the symmetry of the overall motion or with the symmetry of an element of the periodic motion.

Accordingly, we will now introduce a further feature to augment the initial classical particle picture. This will be a supplementary vector quantity specifying a direction in three dimensional space, which we will regard as an axis of symmetry or orientation or polarization direction connected to the periodicity. This will assist us in further characterizing the semi-classical particle whose properties have helped us to develop its associated quantum waves.

Note that with the presence of this newly introduced attribute, the former specifications will no longer fully describe a particle state. Now, in order to fully describe the particle state, we will have to take into account directional sub-states that correspond to the different possible instantaneous directions of orientation associated with the intrinsic periodicity of the particle.

In introducing this new vector quantity, we will not limit ourselves to dealing with the familiar properties of an ordinary classical angular momentum, which is a conserved quantity that does not vary with time for a free particle. Instead, we will find it helpful to consider a directional quantity that does not exhibit the inertia of rotation. Rather, we will instead introduce a new semi-classical quantity that will have a specified instantaneous direction but that may also be capable of undergoing variations in direction, and thus may exhibit changes in directions, perhaps even a large number of times during the course of an observation. Still, such a property would still seem to be related at least indirectly to spin, as an object or particle's spin is a measure of its self-rotation, and is essentially the direction a particle turns along a particular axis. While quantum mechanical spin is treated as a spinor rather than a vector, we will in this semi-classical context work with a directed quantity that is a vector.

**Some parenthetical remarks on the frequency of the periodicity, statistical fluctuations, and further features**

Up until now, we have identified no specific properties in relation to the spatiotemporal function or wave that we have been constructing that would connect directly and explicitly to the statistical or probabilistic features characteristic of quantum mechanics. We did however find a hint in the examination of the particle distribution for a free particle on the basis of special relativity that, while the phase of a quantum mechanical wave function derives directly and immediately from the simple spatiotemporal functions or waves that we have been working with, that in order to obtain the amplitude of the associated quantum waves we may have to develop a superposition of these elementary waves based on an ensemble of such states. This might perhaps correspond to an effective superposition corresponding to an experimental observation over macroscopic time intervals of many such elementary waves.

Now, the possible role of a changing direction of symmetry for periodic behavior might in itself introduce a feature that could provide for rapidly changing input that would be observed in terms of averages during experimental observation, and thus connect directly and explicitly to the statistical or probabilistic features characteristic of quantum mechanics. We continue by examining some potential probabilistic features of this periodic (or quasi-periodic) behavior.

The frequencies associated with particle masses must necessarily be extremely high frequencies so as to reproduce observed quantum behavior. Frequencies must be in the range of those specified by the de Broglie relationship ($f_{dB} = mc^2/h$, where $f_{dB}$ is the de Broglie frequency, m is the particle's mass, c is the speed of light, and h is Planck's constant) or as similarly discussed in other approaches such as a vacuum flux impact model (de Broglie, 1923; Wikipedia(1)). (For example, in the case of an electron, the de Broglie frequency is approximately $10^{20}$ per second.) Because of this, any currently feasible laboratory observations of such a spatiotemporal function or quantum wave associated with a particle must necessarily consist of averages over very large numbers of temporal cycles of the particle's periodicity. This obligatory averaging is an interesting feature which has the potential for, on the one hand, permitting the introduction of statistical fluctuations (thus providing for the statistical and probabilistic manifestations of quantum physics), and also, as a consequence of the averaging of very large numbers of contributions, providing for the possibility of a very precise averaging over any such statistical fluctuations.

**Representation of intrinsic directionality of particle periodicity**

Our semi-classical particle moving through space at uniform velocity is now additionally characterized by an instantaneous direction associated with its intrinsic periodicity. Classically, its orientation could point along any direction in three dimensional space. By introducing directionality, we have thus introduced a whole new set of possible sub-states. Each semi-classical free particle state now can be regarded as having an infinity of different directionally oriented sub-states, that are otherwise equivalent.

So, how might such a polarization or directional characteristic associated with the periodicity of a particle relate to the quantum wave? How could this directionality be expressed in the properties of the wave?

We noted earlier that a wave representing a free particle in uniform constant velocity motion would be representative not only of the initial classical particle that we started with, but also of any other version of that particle located anywhere else in space. Now, somewhat similarly, it appears that we must regard the quantum wave that we are constructing to describe a free particle moving at constant velocity as representing such a particle having any orientation in space, and thus all such directional sub-states should be represented in the quantum wave.

Furthermore, we are allowing for the possibility that while directional axes are specified instantaneously, they may change during the course of a macroscopic measurement, so that actual measurements could involve averages over directional sub-states.

**Preferential orientation and relative orientation for periodic motions**

For the case of an object at rest, we have no reason to suppose that any preferential orientation in space exists, since space as we know it appears to be isotropic. Individual free classical objects seem to exhibit no preferential orientation. Thus, if a particle state is characterized by only a single orientation or polarization, the state could be expected to exhibit properties that are independent of direction (apart from the possibility of some possible correlation with the direction of translational motion of the particle), and thus the various directional sub-states would be expected to be indistinguishable.

However, we need to consider the possibility that a combined state of motion involving two or more orientation or polarization directions might exhibit a dependence on their relative orientation. If so, relative orientation might play a more significant role, and this would be expected to occur in situations in which a semi-classical particle would be treated as existing simultaneously in two or more separate elementary classical states. Could the difference between the directional orientations of different coexisting semi-classical particle states have a role in affecting the quantum wave associated with the particle?

**Representation of multiple simultaneous intrinsic periodic motions in three dimensional space**

Now that we have introduced a directional aspect to periodicity, every elementary free particle state will be regarded as having not only its own characteristic translational motion but also as having its own intrinsic periodic motion. In order to be able to deal with quantum states having translational motions more complicated than simple constant velocity free particle motion, we must be able to deal with the possibility of a particle existing simultaneously in more than a single elementary state of constant velocity translational motion. Each such particle state now has its own associated intrinsic directional periodic motion as well. Therefore, when states of translational motion are combined, it will now be necessary also to combine the associated intrinsic periodic motions. These intrinsic periodic motions may differ with respect to their orientation in space. Thus, while each directional vector characterizing such a state may be regarded as having the same amplitude, it may exhibit a different orientation in space. It would appear that these various different ways of combining different possible vector orientations must be considered as corresponding to the various combined sub-states of the original state of combined translational motion.

**Multiplicity of directionally different and distinguishable particle states**

In order to advance beyond the development of the simplest waves and in particular to further the analysis of bidirectional motion, we will now go beyond a single periodicity and examine a configuration involving two simultaneous intrinsic periodic motions with potentially differing orientations in three-dimensional space. In what various ways can these be present simultaneously so as to permit combination to form a combined state? The answer to this question is determined by vector geometry in three dimensional space. From geometrical considerations in three dimensional space we can evaluate the relative numbers of ways that the associated vectors could combine classically, in terms of the angular separation between them that results from their difference in angular orientation. This will enable us to evaluate a measure of the relative number of classical sub-states associated with a combined state having a particular angular difference in separation between the two intrinsic periodicity axes, as a function of angle.

Suppose that we have two similar vectors that have equal magnitudes and are individually capable of different orientations in three dimensional space. How many different ways could they combine? That would dependent on the opening angle between them. We can evaluate the ratio of the number of ways that they could combine for a particular angular difference in orientation, which is dependent on the relative number of ways that they can point in three dimensional space.

Let's examine how we might evaluate the number of semi-classical sub-states as a function of angle:

Let us consider first the ways that two vectors could combine to form a resultant vector along a particular direction in space. We can visualize the two vectors extending from a common point of origin. Then the resultant vector would also extend from the common point of origin and would be directed midway between the two original vectors. All such combining vectors considered together would form a cone having an axis along the direction of the resultant vector and having an apex angle equal to the opening angle or angle of separation or difference in angular direction between the vectors.

How many sub-states could be formed by these vectors as a function of angle? Let's look at some examples. There would be only one single direction along which these two vectors could point for the case in which the vectors are aligned so that there is a zero degree opening angle or no angular separation between them. But for any other opening angle, there would be a continuum of directions that point along the corresponding cone, with an increasing number of directions available as the conical apex angle increases from zero degrees to somewhat larger angles.

For a given opening angle between the vectors corresponding to the apex angle of the cone (an angle which we will designate as $2\alpha$), the relative number of directions that either vector could point would be proportional to the perimeter of the circle on which the forward ends of the vectors would lie for that opening angle. But the perimeter of such a circle is proportional to the radius of such a circle, and the radius of the circle is in turn is proportional to the sine of the half opening angle (or half apex angle) $\alpha$ characterizing the cone. Thus the relative number of ways of combining these vectors in space would appear to vary as $\sin \alpha$, where $2\alpha$ is the opening angle between the vectors or angle of separation or difference in angular direction between the vectors.

Another method of counting states might be considered, which might provide a different take on the problem:

Starting with one vector pointing along an arbitrary direction, consider a second vector also extending from the common point of origin. The directions that the second vector could point while maintaining an angle $\xi$ to the first vector would form a cone around the first vector with an apex angle $2\xi$. The relative number of directions that that second vector could point while maintaining an angle $\xi$ to the first vector would be proportional to the perimeter of the circle formed by all of the possible direction on which the forward ends of the second vectors would lie when their tail ends coincide with the tail end of the first vector. Again, the perimeter of such a circle is proportional to the radius of such a circle, and the radius of the circle is in turn proportional to the sine of the opening angle ξ between the first and second vectors (and half the angle characterizing the apex angle of the cone). Thus the relative number of ways of combining these vectors in space would appear to vary as sin ξ, where 2ξ is the apex angle of the cone formed by all of the possible orientations of the second vector maintained at an angle ξ to the first vector.

While this second approach come up with a somewhat different result, for both cases the number of sub-states varies as a function of an angle associated with the separation of the directional vectors in three dimensional space. Using either approach, it appears that the relative number of classical directional sub-states (corresponding to independent orientation or polarization directions) associated with a combined state could be expected to depend on the sine of the relevant angle associated with the difference in orientation or polarization directions of the combining sub-states.

We have used a geometrical argument to evaluate the relative number of sub-states associated with the possible orientation directions associated with the intrinsic periodicity, and we find that the relative number of sub-states varies as the sine of an angle related to the difference in orientation of the two combining directional vectors. While these two arguments arrive at differing angles, the expressions that result share the same functional form, so we can use that common functional form. We will designate the relevant angle as θ, and then we can express the fact that the relative number of sub-states will vary with the sine of the angle θ.

Thus, the relative number of ways in which these two directional vectors could combine in three dimensional space as a function of angle would be proportional to the sine of the relevant angle. We can formalize this by writing:

$$N(\theta) \sim \sin \theta \qquad (14)$$

Here, $N(\theta)$ is a measure of the relative number of directional substates associated with the angle θ that enters the relevant argument above. The angle θ provides a measure of the difference in orientation between the directions of symmetry associated with the intrinsic periodicities of the two combining states. Eqn. (14) tells us that the relative number of directional substates for the combined state is proportional to the sine of this angle relating to the angular difference in direction of the orientation or polarization vectors of the two combining periodicities.

Accordingly, we hypothesize that on average the extent of the participation of combined states will vary sinusoidally with a principal angle that plays a role in the analysis.

We note that somewhat similar considerations might also apply to periodic or quasi-periodic motions other than rotational motion, such as, for example, vibrational motions which may change their direction of vibration with time, or segmental stochastic motions in three dimensional space. Thus, sinusoidal variation in the number of participating directional sub-states or multiplicity with angle would seem to be a feature of a fairly general classical description of the directional aspects of periodic behavior.

Note that by the very act of considering the relative number of directional sub-states we are implicitly allowing for a result that will be an average over sub-states. If more semi-classical sub-states are present in association with a given angle, that would be expected to lead to an enhanced likelihood of the particle being observed in circumstances involving that angle.

**Wave magnitude as a measure of the number of directional sub-states associated with the spatial phase of a combined state of motion; and a more detailed examination of spatial periodic behavior in bidirectional motion**

But how could this enhanced likelihood of the particle being observed at certain angles of relative orientation be expressed in the quantum wave?

Let's see what we can learn about how the magnitude of a wave describing the combined state could vary with such a difference in angle. It would appear that the wave magnitude could be proportional to the relative number of particle states or substates that participate as a function of such a difference in angle. The variation in the magnitude of such a wave with angle might be expected to represent the number of participating states or sub-states at the particular value of such a difference in angle of the wave at which the magnitude is evaluated. This would suggest how the characteristics of the corresponding wave might relate to this difference in orientation angle of the semi-classical particle states.

In an earlier section, we examined how the spatial behavior of a standing wave characterizing bidirectional motion depends on the difference between the angular phases associated with the two combining states of motion. We found that the difference function, $\varphi_{diff}$, which corresponded to half the difference between the phase angles, would necessarily be equal at any spatial location to the quantity $(2\pi\gamma x/\lambda)$, as given by Eqn. (13). Thus, the spatial behavior of this standing wave would appear to be governed by the angular difference between the participating states.

Now, in Eqn. (14), we saw that the relative number of sub-states or classical multiplicity associated with a difference in directional orientation between the directions of symmetry associated with the two particle states will vary as the sine of an angle, $\theta$, that is a measure of the difference in orientation direction, and in particular can be identified with half the angular phase difference. Thus, we will consider identifying the angle $\theta$ in Eqn. (14) with the angular phase difference function $\varphi_{dif}$ that appeared in Eqn. (13). Combining these results will enable us to introduce a functional mathematical dependence for determining the relative value of the wave magnitude.

We will set the difference in phase angle of the two combining waves, $\varphi_{dif}$, equal to the angle $\theta$ that measures the difference in orientation between the directions of symmetry associated with the difference in orientation between the directions of symmetry of the two combining states:

$$\theta = \varphi_{dif} \qquad (15)$$

Using Eqn. (14), this enables us to express the relative number of ways of forming sub-states in terms of the spatial angular phase $(2\pi\gamma x/\lambda)$.

$$N(x) \sim \sin(2\pi\gamma x/\lambda) \qquad (16)$$

Thus, combining the relevant information from the different aspects of the problem, we find that we may expect that the difference in phase between the two waves will depend not only on a spatial parameter and exhibit a spatial periodicity, but also that the net wave will exhibit a magnitude that varies as the sine of a phase angle. Taken together, these effects of periodicity on the spatial behavior of the wave are now encoded as the phase and functional form with the relative magnitude of the net wave characterizing bidirectional motion.

Thus we are led to the conclusion that the spatial behavior of a standing wave describing bidirectional motion would include a functional form whose magnitude reflects the number of participating directional sub-states, and hence would be expected to have the mathematical form:

$$f_s(x) = A \sin 2\pi\gamma x/\lambda \qquad (17)$$

Here A is a constant amplitude factor. The wavelength $\lambda$ characterizing the wave corresponds to the common periodicity of the two combining states that describe the back-and-forth translational motion. The sinusoidal magnitude of the standing wave reflects the contribution provided by the relative number of sub-states expressed as a function of the spatial coordinate x.

We note that Eqn. (17) can be recognized as identical in form to the spatial part of the usual quantum wave function describing back-and-forth motion. Our results tell us that the spatial behavior of the wave describing back-and-forth motion may be regarded as mirroring an average over angular differences in periodic behavior of the two participating semi-classical particle states.

We note that these results indicate that the magnitude of the wave for bidirectional motion appears to originate from a fundamental characteristic of vector behavior in three dimensional space. The relative number of directions and hence the relative number of directional sub-states will be given as a function of angle by the sine function, and it appears that it is this that leads to the sinusoidal dependence of the spatial magnitude of the combined quantum waves in bidirectional motion.

In this formulation, the statistical character of quantum mechanics would appear to arise in association with a rapid variation in the direction of the spatial periodicity, thus bringing into play the various directional sub-states. It seems likely that the variation in direction in three dimensional space occurs at a rate comparable to the de Broglie frequency itself, although that remains to be established by further analysis.

To summarize: we have been developing a wave representing bidirectional or back-and-forth motion on the basis of combining two individual elementary free particle waves moving along opposite directions, and found disjoint temporal and spatial dependence representing a standing wave. Furthermore, as a consequence of the introduction of an orientation or polarization of the particle leading to associated classical directional sub-states, we have identified an explicit sinusoidal dependence of the quantum wave magnitude on the spatial phase.

**Developing an explicit mathematical description for free particle constant velocity motion**

Now that we have developed an explicit mathematical expression for the spatial part of the standing wave representing bidirectional motion, we will examine how it can be decomposed into oppositely directed component waves.

We found that the spatial behavior of the bidirectional standing wave will exhibit sinusoidal behavior described explicitly by Eqn. (17).

A sine function can be expressed as a combination of two complex exponential functions, as is expressed in the mathematical identity that $\sin y = (e^{iy} - e^{-iy})/2i$. Using this mathematical identity, we can reexpress the spatial part of the standing wave expressed in Eqn. (17) as follows:

$$f_s(x) = A \sin(2\pi\gamma x/\lambda) = (A/2i)e^{2\pi i\gamma x/\lambda} - (A/2i)e^{-2\pi i\gamma x/\lambda} \tag{18}$$

Thus, the sine function characterizing the spatial part of the standing wave that represents bidirectional motion can be separated into components that would characterize the spatial behavior of two oppositely directed running waves of the form $e^{2\pi i x/\lambda}$ and $e^{-2\pi i x/\lambda}$.

We have already seen in Eqn. (9) and Eqn. (10) that the two running waves must exhibit phases $2\pi\gamma(ft - x/\lambda)$ and $2\pi\gamma(ft + x/\lambda)$ respectively. Since the spatial portion of these waves must obey Eqn. (18), it follows that the temporal portion of both of these waves must be an exponential of the form $e^{i2\pi\gamma ft}$, so that these running waves will individually have the forms:

$$f_+(x) = -(A/2i)e^{i2\pi\gamma(ft-x/\lambda)} \tag{19}$$

and

$$f_-(x) = (A/2i)e^{i2\pi\gamma(ft+x/\lambda)} \tag{20}$$

Thus, we are led to complex functions as descriptive of the simplest quantum waves.

Eqn. (19) and Eqn. (20) are explicit representations of running waves. We also note that the mathematical expressions for these running waves in Eqn. (19) and (20) have fixed amplitudes, and that these waves do not vary in magnitude but only in phase. Accordingly, they describe particles uniformly distributed in space, but at different phases of their periodicities at different spatial locations.

Furthermore, combining the two full running waves specified in Eqn. (19) and Eqn. (20) to form a standing wave shows that the full time dependent quantum wave describing the back-and-forth motion must then have the complete form:

$$f(x) = Ae^{i2\pi\gamma ft}\sin(2\pi\gamma x/\lambda) \tag{21}$$

So we also see from this that the back-and-forth motion must have exactly the same mathematical behavior in time dependence as do the running waves.

These results indicate that the quantum waves that provide explicit mathematical descriptions of constant velocity motion or bidirectional motion respectively can therefore be seen to be determined by and to originate plausibly from the way that the periodicities associated with different states of the same particle must combine.

**Introducing forces**

We have examined with some success how quantum waves associated with free particles can originate.

So, next we will consider what could happen when a particle is no longer free but is instead subjected to forces and will therefore be accelerated and hence must engage in variable velocity motion. What sorts of associated quantum waves might we expect under those conditions?

It would seem reasonable to attempt to generalize the free particle cases to cases of accelerated motion by continuing our approach of finding a spatiotemporal function exhibiting spatial periodic behavior that depends on the velocity of the particle, and modify it to accommodate the possibility of varying velocities at different positions in space. We could thus attempt to generalize to waves with wavelengths dependent on the local velocity of the classical particle, and therefore dependent on the local value of the potential associated with the forces acting on the particle.

As before, we could again introduce a periodicity with a frequency proportional to the mass or total energy of the particle. With these assumptions, the mass or total energy of the classical particle in the classical field would determine the temporal periodicity (frequency) of the wave. If the force field is conservative so that the total energy of the particle remains constant during the motion, we might then expect a dependence on the time in the form of a complex exponential function with imaginary temporal phase, such as we have just found for free particles in Eqn. (19), Eqn. (20), and Eqn. (21). Furthermore, with these same assumptions, the kinetic energy (corresponding to the difference between the total energy excluding the rest mass energy, and the potential energy at any point) would determine the local velocity and hence the local spatial periodicity or wavelength of the wave.

> In terms of the parameters governing the classical particle, the velocity can be expressed in terms of the kinetic energy K as $v^2 = 2K/m$ or in terms of the total energy and the potential energy, as $v^2 = 2(E - V)/m$, where in this case the total energy E excludes the rest energy. If we have a physical situation in which the potential and hence the particle's potential energy are functions of spatial location, this would enable the introduction of the velocity as a point function of x, as $v^2 = 2(E - V(x))/m$. The assumption that the corresponding wave might be related to the simpler case with the wave dependent on the local velocity would provide a means of bootstrapping up from the simpler case.

We have seen that in the case of constant speed bidirectional motion, the magnitude of the wave can be described by a sinusoidal function, while for constant velocity motion, the free particle waves are described by complex exponential functions with imaginary exponents. How could we attempt to generalize this behavior? Sinusoidal functions are trigonometric or circular functions that are usually defined in a geometrical context. However, it is also possible in effect to define sinusoidal behavior in the context of differential equations. Such a defining differential equation would determine the waves associated with the classical particle as the solutions to the differential equation.

> Let's examine how it is possible to specify or define sinusoidal behavior in the context of differential equations. Essentially, a sinusoidal function may be defined as a solution to a differential equation of the form:
>
> $$d^2 f(x)/dx^2 = a\, f(x) \qquad (22)$$
>
> Here, the coefficient a is a constant, because the second derivative of a sinusoidal function is just a constant multiple of itself.
>
> It would seem to make sense to start with such a defining differential equation and generalize it so as to develop a differential equation to govern waves associated with the more general behavior of particles subject to forces.
>
> The exponential functions have the property that their derivatives are constant multiples of the original functions, and therefore their second derivatives are also constant multiples of the original functions. Accordingly, similar considerations will apply to the complex exponential functions describing free particle waves in Eqn. (19) and Eqn. (20). Thus, all of the waves that we have obtained so far are solutions to Eqn. (22), the simple differential equation given above that basically defines a sinusoidal function. (Partial derivatives would be applied for circumstances in which variations in both space and time are expressed.)

What would the coefficient a be equal to in these cases? In the case of constant speed bidirectional motion, we found from Eqn. (17) and Eqn. (20) that the function $f(x)$ in its spatial dependence would have the form $f(x) = A \sin(2\pi\gamma x/\lambda)$. This function describing the spatial behavior of the wave for back-and-forth motion must then satisfy the differential equation defining sinusoidal motion, Eqn (22). For it to do so, the coefficient a must equal $(-4\pi^2\gamma^2/\lambda^2)$. In terms of the velocity, the coeffcient a would be expressed as $(-4\pi^2\gamma^2 m^2 v^2/b^2)$, or in terms of the kinetic energy K the coefficient a would be expressed as $(-8\pi^2\gamma^2 mK/b^2)$. For the non-relativistic case, we can write for the coefficient $a = -8\pi^2\gamma^2 mK/b^2$.

Next, we will consider generalizing Eqn. (22), a differential equation that with a suitable constant coefficient governs free particle wave behavior, to attempt to describe more general wave behavior accommodating physical conditions requiring variable particle velocities. We can go about this by allowing the coefficient a to vary as a function of the parameter x, as $a(x)$. We will explore the possibility that this approach might enable us to develop a differential equation for determining wave forms in more general cases of particle motion under circumstances where forces are present.

Accordingly, we can consider applying as a requirement imposed on a quantum wave accompanying a particle experiencing a force field, that it be governed by the more general differential equation:

$$d^2f(x)/dx^2 = a(x) f(x) \qquad (23)$$

Here, the coefficient $a(x)$ is no longer a constant but can vary throughout space to accommodate the fact that the classical particle velocity must vary depending on the potential. Generalizing from the earlier expression for the constant coefficient a in the differential equation, we could now express the variable coefficient as $a(x) = (8\pi^2 m/b^2)K(x)$. If we express the kinetic energy $K(x)$ in terms of the total energy E (omitting the mass energy) and the potential energy $V(x)$, as $K(x) = E - V(x)$, and insert this information into Eqn. (23), we find as a governing equation:

$$d^2f(x)/dx^2 - (8\pi^2 m/b^2)(E - V(x)) f(x) = 0 \qquad (24)$$

This equation, Eqn. (24), has the form of the Schrödinger equation that governs the behavior of wave functions in quantum mechanics. If we identify the so far unspecified constant b with the Planck constant h, then this equation, Eqn. (24), can be seen to correspond precisely to the time-independent Schrödinger equation for a particle in a conservative force field.

Thus we see that the time-independent Schrödinger equation that governs quantum mechanical wave functions can be arrived at using a quite straightforward approach based on elementary considerations.

**Discussion and closing remarks**

These results provide evidence that both special relativity and periodicity are fundamental to quantum mechanics. In some sense, quantum mechanics seems to be all about periodicity, including its directional characteristics. Quantum mechanics thus would seem to be telling us about periodic behavior in time and space seen in different frames of reference.

We have been able to use a simple and direct approach to developing quantum wave functions in a relatively straightforward manner, starting from consideration of a classical particle in the context of special relativity. First we effectively synchronized clocks in the rest frame of the particle so that this time coordinate in the particle's frame of reference would be the same throughout all space. Then we examined this time coordinate in a different frame of reference, the laboratory frame of reference. The time coordinate served as our initial candidate spatiotemporal function to use in the development of a quantum wave associated with the particle.

We then introduced a periodicity in association with the classical particle, following the conceptual approach of de Broglie, and based on the de Broglie postulate that every particle with mass necessarily exhibits a temporal periodicity. Consideration of the wave phase then led to a temporally periodic oscillation that is in phase throughout all of space in the rest frame of the particle, in agreement with the quantum wave function for a stationary particle. When we examine the time coordinate or associated phase in any other frame of reference (e.g. the laboratory frame) that is moving at constant velocity relative to the rest frame of the particle, the associated wave appears as a running wave. The waves that we introduced initially are elementary waves that carry phase information only, and we speculated that a combination of these elementary waves to represent an ensemble of many different cases could contribute to forming full quantum wave functions having magnitudes associated with the particle distribution in space.

Relying on special relativity, we found a many-to-one relationship between localized classical particle states and the wave associated with constant velocity particle motion, thus bringing in the quantum concept of a particle being simultaneously present in multiple locations. Thus, a quantum wave representing a free particle in uniform constant velocity motion would be representative not only of the initial classical particle that we started with, but also of any other version of that particle located anywhere else in space.

We found it of value to work with the idea that different classical particle states might coexist simultaneously under circumstances in which quantum waves would form. By introducing and working with an average time parameter for the different classical states, we were enabled to introduce and examine a candidate wave structure for bidirectional motion that described a standing wave.

We introduced the idea that directionality should be associated with the intrinsic periodicity that, in accordance with de Broglie's hypothesis, must be present for any particle. As a result, it appears that we must regard the quantum wave that we are constructing to describe a free particle moving at constant velocity as representing such a particle having any orientation in space, and thus all such directional sub-states should be represented in the quantum wave.

When classical free particle states are present in combination, these intrinsic directional sub-states will also combine. We found that the relative number of these directional sub-states depends on the difference in orientation and thus on an angular difference, and that this would result in spatially varying wave magnitudes as a function of angular phase. The magnitude and interference of wave functions in quantum mechanics thus seems to relate to the differing phases in the periodicity of participating semi-classical particle states used in describing a quantum particle. Thus, perhaps surprisingly, wave magnitudes in ordinary quantum mechanics appear to depend on the directional features of the periodicity.

We also permitted this directionality associated with the intrinsic periodicity to change with time. It appears possible that stochastic changes in the direction associated with the periodicity could be regarded as the mechanism that introduces the statistical nature of quantum mechanics into a semi-classical picture.

While we found that the wave phase characterizes each individual elementary spatiotemporal wave, we found also that the wave magnitude appears to originate instead as an average property over many elementary waves and many semi-classical directional sub-states and corresponds to observations that would sample many individual periods of the intrinsic periodicity. Thus, an ordinary quantum wave might itself be regarded as a time-averaged description of sub-quantum phenomena, in that the elementary waves and relative number of participating directional sub-states would determine the magnitude of the net quantum wave function.

We have based the analysis of the simplest cases on use of the relativistic time coordinate of the particle's frame of reference as measured in the laboratory frame of reference to introduce a spatiotemporal function leading toward a wave function. This is a suitable approach for constant velocity motion because in special relativity, Lorentz transformations relate the different frames of reference. However, in the more general cases of arbitrary motion or combined motions, time is not well defined, and it is not possible to assign a unique value of time throughout an arbitrarily moving frame of reference or combined frames of reference. Hence that method in its simplest form could not be followed. But we have seen that it was still possible to introduce a related approach, the use of an average time, that seems to be of value for a physical description of bidirectional motion. A generalization of this approach might lead to steps for forming linear combinations of elementary free particle waves to describe more complicated behavior.

We addressed the case of a particle in a conservative force field by generalizing from simpler behavior already studied in order to develop a differential equation dependent on the local classical particle velocity, and this differential equation turned out to have the functional form of the Schrödinger equation.

Since special relativity has been fundamental to developing these quantum wave functions, we suggest that further examination might potentially provide additional insight into and connections between the properties of particles and the characteristics of space-time. Similarly, a better understanding of the role of periodicity might further clarify the relationship between quantum mechanics and classical mechanics.

We hope that readers will find a certain satisfaction in being able to experience the development of some significant features of quantum mechanics by starting from classical particle properties, and that this approach will contribute to making the study of quantum mechanics both more intuitive and enjoyable, and possibly also more insightful.

interpqvfishortnew.doc
3 December 2008 draft, revised